**Generative AI Spotlights the Human Core of Data Science: Implications for Education**

Nathan Taback[1]
[1] Department of Statistical Sciences, University of Toronto, Toronto, Ontario, Canada

**Abstract**

Generative AI (GAI) reveals an irreducible human core at the center of data science that is counterintuitive: advances in GAI should sharpen rather than diminish the focus on human reasoning in data science education. GAI is now capable of executing many routine data science workflows including cleaning, summarizing, visualizing, modeling, and drafting reports, yet the competencies that matter most remain irreducibly human—the human core: problem formulation, measurement and design, causal identification, statistical and computational reasoning, ethics and accountability, and sensemaking. Drawing on Donoho's Greater Data Science framework, Nolan and Temple Lang's vision of computational literacy, and the McLuhan-Culkin insight that we shape our tools and thereafter our tools shape us, I trace the emergence of data science through three converging lineages: Tukey's intellectual vision of data analysis as a science, the commercial logic of surveillance capitalism that created industrial demand for data scientists, and the academic programs that followed. Mapping GAI's impact onto Donoho's six divisions of Greater Data Science reveals that computing with data (GDS3) has been substantially automated, the divisions Donoho identified as most important, data gathering, preparation, and exploration (GDS1) and science about data science (GDS6), require human input. The educational implications are that data science curriculum should focus on the human core while understanding how human input is essential in GAI as part of iterative prompt–output–prompt (POP) cycles with retrieval-augmented generation and learning outcomes and assessments should include and evaluate reasoning and judgment.

# 1. Data Science at an Inflection Point

### 1.1. The Medium Is the Message

Marshall McLuhan's dictum, "the medium is the message," reminds us that technologies reshape how we think and learn, often in ways that have little to do with the content transmitted through them. A new medium does not simply deliver old content more efficiently; it restructures the cognitive environment in which people work. As McLuhan's colleague John Culkin put it, "We shape our tools and thereafter our tools shape us" (Culkin, 1967).

McLuhan (1964) argued that the form of a medium has a more significant impact on society than the content it carries: "the content of any medium is always another medium." When a new medium arrives, we often do not yet understand its unique message; instead, we use it to deliver the content of the old medium until the new one finds its own language. The content of print is the written word; the web browser incorporates all previous forms (text, audio, video) as its content (McLuhan, 1964; McLuhan & Fiore, 1967).

In the past decade the medium that data science operates in could loosely be described as the modern data science stack—programmable software that include tools for data ingestion, storage, transformation, statistical analysis and machine learning. Generative AI (GAI) is a new medium—it is a medium that uses these technologies as inputs. Large language models are now capable of executing much of the routine data science workflow: cleaning and reshaping datasets, writing and debugging code in multiple languages, fitting statistical and machine learning models, generating visualizations, interpreting outputs, and drafting reports. We are still grappling to understand this medium's new messages, yet we are currently living with this new medium. *What should students still learn* given that the standard data science workflow can now be automated? *What, if anything, is irreducibly human about data science*?

These questions create a familiar novice dilemma. If students rely on GAI before they understand a concept, they cannot judge the output. The medium shapes their reasoning before they have developed the knowledge needed to resist its distortions. Just as a calculator returning 29 for 12×17 is obviously wrong only to someone who knows arithmetic, a GAI system producing a plausible but misleading analysis is detectable only by someone who understands the underlying statistical reasoning, the data-generating process, or the domain context. GAI can extend our capabilities only if we have capabilities to extend.

Data science has an irreducible human core: a set of competencies that remain essential regardless of the medium through which data science is practiced. These competencies include problem formulation, measurement and design, causal identification, statistical and computational reasoning, ethics and accountability, and sensemaking. GAI does not diminish the importance of these competencies; it exposes them as the center of the discipline by automating the data science stack that previously surrounded and sometimes obscured them. Mollick (2024) has persuasively argued that AI should be understood as a *co-intelligence* and that humans must remain in the loop via his "Rule 2" for working with AI. This paper takes that imperative as a starting point and asks: what, specifically, should the human in the data science loop be *doing*? The answer is the human core. The educational implication is direct: these competencies should define the learning outcomes around which data science programs, curriculum, instruction, and assessment are designed.

Several recent contributions have explored how GAI changes data science education (Tu et al., 2024; Lopez-Miranda et al., 2026). This paper contributes to that conversation by offering a systematic mapping of GAI's impact onto Donoho's GDS framework, situating the educational question within the commercial logic of surveillance capitalism, and framing GAI as a *medium* in the McLuhan sense rather than merely a tool — each of which yields different insights than treating GAI as a pedagogical challenge alone.

To understand why this moment feels like an inflection point, it helps to trace how the field arrived here. Data science as it exists today emerged from three converging lineages: an intellectual tradition within statistics that has argued for over sixty years that the field must be broader than mathematical theory; a commercial logic (what Zuboff (2019) calls surveillance capitalism) that created massive industrial demand for people who could extract patterns from behavioral data; and an academic response that tried, with mixed success, to institutionalize the resulting hybrid discipline. GAI disrupts all three lineages simultaneously. The sections that follow trace each in turn before asking: in this new medium, what must students still learn?

## 1.2. The Intellectual Lineage: From Tukey to Donoho

More than sixty years ago, John Tukey published a deeply unconventional article in *The Annals of Mathematical Statistics*. "The Future of Data Analysis" (Tukey, 1962) appeared in a journal whose other articles presented definitions, theorems, and proofs. Tukey's contribution was instead a kind of professional confession. He argued that what he cared about, data analysis, was not a branch of mathematics but an empirical science, one "defined by a ubiquitous problem rather than by a concrete subject." Data analysis, he wrote, was at least as complex as biology, and formal statistical theory could play only a partial role in its progress.

Tukey identified four driving forces acting on this new science: the formal theories of statistics, accelerating developments in computing, the challenge of ever-larger bodies of data, and the emphasis on quantification across an ever-wider array of disciplines. This list is strikingly modern. Every contemporary press release announcing a data science initiative cites some version of these same forces. What was radical in 1962 was Tukey's insistence that formal theory was only one among four, and perhaps not the most important.

Tukey also worried about how data analysis was taught. He observed that statistics tended to be taught as a branch of mathematics, that students had limited exposure to actual data analysis, and that the number of years of contact with practicing professionals was far less for statistics students than for students in the physical sciences. He proposed that teaching data analysis was more like teaching biochemistry — a field overflowing with detailed knowledge that textbooks needed to relate, than like teaching mathematics, where one progresses through proofs. He warned against the view that "avoidance of cookbookery and growth of understanding come only by mathematical treatment, with emphasis upon proofs."

Donoho (2017) synthesized over three decades of such exhortations — from Chambers (1993), Cleveland (2001), Wu (1997), and others who urged academic statistics to expand its boundaries, in "50 Years of Data Science," published in the *Journal of Computational and Graphical Statistics*. Writing on the occasion of the Tukey centennial, Donoho distinguished between what he called "lesser data science," the consensus version taught in newly minted master's programs, amounting to statistics plus some computer science, and "greater data science" (GDS), a substantially larger intellectual field. He organized GDS into six divisions: data gathering, preparation, and exploration (GDS1); data representation and transformation (GDS2); computing with data (GDS3); data modeling (GDS4); data visualization and presentation (GDS5); and science about data science (GDS6). Donoho argued that GDS1 — the messy, judgment-laden work of acquiring, cleaning, and exploring data — consumed the majority of practitioner effort yet was poorly represented in curricula, while GDS4 (data modeling) dominated both teaching and research. He also argued that GDS6, the evidence-based study of how data analysis is actually done and how it might be improved, was the most forward-looking division — the one that could make data science a genuine science in Tukey's sense.

The intellectual lineage from Tukey through Donoho was calling, with increasing specificity, for exactly the kind of broadening that GAI now forces. Tukey emphasized that data analysis "must look to a heavy emphasis on judgement" (Tukey, 1962). When GAI absorbs much of the data science stack that data science programs have spent the last decade teaching, what remains is precisely the territory these thinkers identified as most important: the judgment-laden, context-dependent, human-facing work that formal theory and technical skill were always supposed to serve.

## 1.3. The Commercial Lineage: Surveillance Capitalism and the Data Scientist

The intellectual lineage tells us about the academic forces that have shaped data science. A parallel commercial lineage tells us what it *became*, and why.

In the early 2000s, Google confronted a problem: how to monetize a search engine that was rapidly becoming the most visited site on the internet. The answer, as Zuboff (2019) has documented, lay in a discovery about data. Google's engineers found that the behavioral traces users left behind (search logs, click patterns, dwell times) — constituted a surplus that exceeded what was needed to improve the search product itself. This behavioral surplus could be translated into prediction products: computations that anticipated what a user would click on, look for, or buy. Hal Varian, Google's chief economist, clarified the resulting business model with a candid observation: "All of a sudden, we realized we were in the auction business" (Zuboff, 2019). Advertisers would pay to place their messages in front of users whose behavioral data predicted receptivity.

The mechanism that emerged can be described in straightforward terms. Data science operationalizes the ability for companies to claim private and human experience as mostly free raw material for translation into behavioral data. These data serve as inputs into computational processes — the "machine intelligence" layer, that produce prediction products anticipating what individuals will do now, soon, or later. The prediction products are then traded in what Zuboff calls behavioral futures markets. This economic logic spread rapidly beyond the technology sector. As Zuboff (2019) observes, it now operates across insurance, automobiles, health, education, finance, and virtually every product described as "smart" or service described as "personalized."

This commercial logic shaped the data science profession in specific ways. The "data scientist" emerged as what Dorschel (2021) calls a hybrid professional, simultaneously generalist and specialist, technician and communicator, data exploiter and data ethicist. In 2009, Varian predicted that "the ability to take data — to be able to understand it, to process it, to extract value from it, to visualize it, to communicate it — is going to be a hugely important skill in the next decades" (Varian, 2009). He added, pointedly, that what was scarce and expensive was "the talent to analyze data and make it tell its story." By 2012, the term "data scientist" had displaced "statistician" in popular discourse, and the Harvard Business Review had declared it the sexiest job of the 21st century (Davenport & Patil, 2012).

## 1.4. The Academic Response and Its Limits

The convergence of these two lineages, the intellectual call for a broader field and the commercial demand for data-skilled professionals — produced, beginning around 2015, a wave of academic data science initiatives. Major universities launched multimillion-dollar programs, new master's degrees, and campus-wide centers at a remarkable pace. As Donoho (2017) noted, these announcements were striking to academic statisticians, many of whom felt that the activities being touted as new were activities they had been pursuing for their entire careers.

Donoho offered a sharp diagnosis: most courses in these new programs covered topics a statistics department could or should already teach. The distinguishing feature was the addition of information technology infrastructure (viz. parts of the data science stack)— databases, scaling tools, and production deployment. The result was a compromise that helped administrators launch programs quickly but provided no guidance about long-term intellectual direction. Donoho proposed Greater Data Science an intellectually larger field organized around six divisions, each representing a genuine area of research and teaching. The framework acknowledged that computing with data and data representation were legitimate divisions alongside data modeling, but also insisted that data gathering, preparation, and exploration (GDS1) and science about data science (GDS6) deserved far more attention than they received. One of Donoho's most prescient observations concerned the rapid obsolescence of specific technical tools. He used Hadoop—a variant of MapReduce for distributed computing—as his central example of the "skills" that data science programs claimed to teach. By the time the article appeared, the data engineering community had already moved from Hadoop to Apache Spark, making much of the Hadoop-specific instruction obsolete. Donoho drew the right conclusion: "The rapid obsolescence of specific tools in the data engineering stack suggests that today, the academic community can best focus on teaching broad principles — 'science' rather than 'engineering.'"

Generative AI is different than a new technical tool making another obsolete since it subsumes all these technical tools. The computational skills that data science programs spent a decade building into their curricula—writing code in R or Python, managing databases, building data pipelines, debugging scripts—are now within the capability envelope of widely available large language models. The medium through which data science is practiced has shifted again. The question is not whether GAI changes data science (it manifestly does) but *what* it reveals about the discipline's essential structure. The remainder of this paper argues that what it reveals is a human core: a set of competencies centered on judgment, context, design, and accountability that resist automation not because they are technically difficult but because they are fundamentally about understanding the world that data comes from. The implications for how we teach follow from this disciplinary claim.

# 2. Greater Data Science Meets Generative AI

A systematic examination of how GAI transforms each division of Donoho's Greater Data Science framework reveals which parts of the field have been substantially automated, which remain essentially human, and where the boundary between the two is most consequential for curriculum design.

## 2.1. GAI Capability: Convergence Across Systems

Before mapping GAI onto the six divisions, it is worth establishing what current systems are capable of and, equally important, what pattern their capabilities exhibit across systems. Published benchmarks for code generation and data science tasks — including HumanEval (Chen et al., 2021), DS-1000 (Lai et al., 2023), and BigCodeBench (Zhuo et al., 2024), show that the leading GAI systems perform at broadly comparable levels on standard programming and data analysis tasks. The differences between major systems, while real, are not large relative to the gap between any of them and the state of the art just a few years earlier. For practical purposes, a well-prompted interaction with any leading system will produce competent code for routine data science operations: data cleaning, reshaping, visualization, model fitting, and report generation.

A deeper observation concerns convergence. For a fixed dataset and a well-specified computation, different GAI systems will converge on the same answer. In one sense, this is trivially guaranteed: ordinary least squares regression should yield the same coefficients whether computed in R, Python, or Julia, and regardless of which package or GAI system writes the code. The mathematical determinacy of the computation ensures agreement on results.
But the convergence extends beyond results to *approaches*. When asked to perform the same data science task, different large language models tend to produce similar code: similar package choices, similar idioms, similar analytic strategies. Part of this convergence likely reflects the fact that there *are* conventional best practices for standard tasks, and any competent agent—human or AI—would converge on them. But I conjecture that the convergence also reflects, in part, the shared reinforcement learning pipelines (particularly reinforcement learning from human feedback, or RLHF) through which these systems are trained. RLHF rewards outputs that align with human preferences, and those preferences are themselves shaped by a common body of tutorials, textbook conventions, and community practices (e.g., Stack Overflow answers, widely used vignettes). The result is that different systems, trained on overlapping corpora and tuned toward overlapping preference distributions, converge on what might be called the conventional wisdom of data science practice.

This convergence has an important implication for education. If GAI systems reliably produce the conventional approach to standard tasks, then the places where human judgment matters most are precisely where the conventional approach breaks down: unusual data structures, novel research questions, violated assumptions, or situations where the standard pipeline produces misleading results. The human core of data science lies not in executing conventional analyses, since GAI does that well, but in recognizing when conventionality is inadequate.

A natural objection is that GAI capabilities are improving rapidly, and that competencies currently beyond GAI's reach may not remain so for long. This is a legitimate concern, and the mapping that follows should be read as a snapshot of the present rather than a permanent boundary. However, the argument does not depend on GAI capabilities remaining static. The competencies identified as the human core reflect the structure of learning from data about the world: the need to formulate questions, assess evidence, reason about causation, and take responsibility for consequences. These needs arise from the relationship between data and the world, not from the limitations of any particular tool. Even as GAI improves, these competencies will define what it means to practice data science responsibly. Whether the human exercises them alone or in collaboration with increasingly capable AI systems, the competencies themselves do not disappear.

## 2.2. Mapping GAI onto the Six Divisions of Greater Data Science

I now take each of Donoho's six divisions in turn and assess the degree to which GAI transforms the activities within it. The result is not a uniform picture. Some divisions are substantially automated; others are barely touched; and the pattern reveals something about what was always foundational.

### 2.2.1. GDS1: Data Gathering, Preparation, and Exploration — The Most Resilient Division

Donoho argued that GDS1 is more important than data modeling (GDS4), as measured by the time practitioners spend on it—the commonly cited estimate is that 80% of effort in a data science project goes to gathering, cleaning, and exploring data. Crucially, GDS1 is not defined by specific coding or programming tools. It is about general data collection, preparation, and exploration—activities that require understanding why data exists and what process generated it. GAI can support these activities effectively. A practitioner can prompt a language model to suggest exploration strategies, generate cleaning scripts, flag potential anomalies in a dataset, or propose sanity checks. These are genuine contributions that accelerate the workflow.

But GAI cannot originate the judgment calls that define GDS1. What data should be collected? What counts as an anomaly versus a legitimate extreme value? What is the relevant population? What confounders might lurk in the data-generating process? What measurement conventions does the source domain use, and do they align with the research question? These decisions require domain knowledge, contextual understanding, and the kind of problem formulation that is, as I argue in Section 3, an irreducible human competency. A student who does not understand the domain will be challenged to evaluate whether GAI's suggestions for data exploration are on target or beside the point.

GDS1 is the most resilient division under GAI because it was never primarily about technical execution. It was always about judgment exercised in context, precisely the territory that GAI cannot currently occupy.

### 2.2.2. GDS2: Data Representation and Transformation — Partially Automated

GAI is highly competent at the mechanical aspects of data representation: writing code to reshape data frames, converting between file formats, generating SQL queries, building ETL pipelines. A student who needs to pivot a dataset from wide to long format, or merge tables on a key, can prompt a language model and receive working code in seconds. What remains human is the judgment about which representation is appropriate for a given analysis. Understanding why tidy data principles matter (Wickham, 2014)—not just how to execute a pivot_longer() call—requires grasping the relationship between data structure and analytical operations. Similarly, choosing appropriate mathematical representations for specialized data types (spectral transforms for acoustic data, wavelet representations for signals, graph structures for network data) requires understanding when and why these representations reveal structure that the raw data obscures. GAI can implement any of these transformations on request, but the request itself presupposes knowledge that the student must bring.

### 2.2.3. GDS3: Computing with Data — Substantially Transformed

This is the division most dramatically affected by GAI, and examining it closely illuminates how the medium of data science has shifted.

Donoho described GDS3 as encompassing knowledge of multiple programming languages, understanding of computational efficiency, development of workflows and packages, and facility with cluster and cloud computing. Nolan and Temple Lang (2010), in an influential article on computing in the statistics curriculum, made the case that computational literacy and programming are as fundamental to statistical practice and research as mathematics. They proposed a comprehensive set of topics for a 15-week undergraduate course: language syntax, data structures, control flow, regular expressions, shell tools, relational databases and SQL, simulation, debugging and profiling, text processing, version control, and report generation. Their argument was exactly right for 2010, and Donoho's GDS3 reaffirmed it in 2017.
But GAI is now capable of performing most of the tasks on Nolan and Temple Lang's syllabus. It can write regular expressions, compose and optimize SQL queries, parse and clean messy text data, generate publication-quality visualizations, build and document R packages, debug code, and produce reproducible reports. Placing their Table 1 topics alongside current GAI capabilities yields a striking picture: the specific technical skills they identified as requiring deep student investment have been substantially absorbed into GAI's capability envelope. This does not make their argument wrong. It makes it historically situated. The medium changed.

Around ten years ago at the University of Toronto, the Department of Statistical Sciences undertook a comprehensive renewal of its undergraduate program, that included introducing program learning outcomes related to GDS3 such as computer simulation to introduce inference and probability as entry points into programming in R and Python. The pedagogical logic was sound: students learn statistical concepts by implementing them computationally, and they learn programming by working with data in a meaningful context. But GAI complicates this logic in a specific way. If a student can prompt a language model to produce a simulation study of the central limit theorem, complete with visualizations and interpretation, in seconds or even generate an interactive web-based demonstration on the fly, what is the purpose of having the student write the code themselves? In my opinion the purpose was never really the code. Nolan and Temple Lang's deeper goal was computational *reasoning*: the ability to think abstractly about computational problems, to extract concepts from a specific implementation and transfer them across languages and environments. They warned against the "do-it-yourself 'lite' approach" of teaching programming through crash courses, arguing that it sent a signal that the material lacked intellectual importance and left students with bad habits and shallow understanding. The reasoning they advocated included understanding why a particular data structure is appropriate, what a function abstraction accomplishes, how to validate code through diagnostic statistics on output. What has changed is the vehicle through which that reasoning is expressed. When the medium included R or Python running locally on a laptop, coding was the vehicle for computational thinking. When the medium is GAI, prompting, evaluating, and iterating become the vehicle. The reasoning endures; its expression changes.

This is McLuhan's insight that the medium is the message. The medium shapes how students encounter computational ideas (messages). In the R/Python medium, students who wrote a for loop to simulate sampling distributions were simultaneously learning about sampling variability and control flow. In the GAI medium, students who prompt a system to produce the same simulation and then interrogate the output (testing edge cases, modifying assumptions, checking whether the code does what it claims) are learning about sampling variability through a different cognitive process. The educational challenge is to ensure that the GAI medium develops genuine reasoning rather than fluent prompting without understanding, which is the novice dilemma identified in Section 1.

### 2.2.4. GDS4: Data Visualization and Presentation — A Mixed Case

GAI can generate standard statistical graphics such as histograms, scatterplots, time series plots, heatmaps, with minimal prompting, and can produce more elaborate visualizations when given detailed instructions. Dashboard construction, once a specialized skill, is increasingly within GAI's reach.

What remains human is what I will call *sensemaking* (Weick, 1995): the broader activity of translating analytical results into understanding that is actionable by specific audiences. Sensemaking in this context encompasses more than communication in the narrow sense of producing clear charts and well-written reports. It includes identifying who the stakeholders are and what decisions the analysis is meant to inform; judging what to show and what to leave out; framing uncertainty in terms that a non-technical audience can reason with; anticipating how results might be misinterpreted; and engaging collaborators and decision-makers in interpreting findings rather than simply delivering them. Sensemaking, so defined, is a relational and contextual activity. It requires understanding not just the data but the institutional and human environment in which the data will be used. GAI can draft a summary paragraph or suggest a chart type, but it cannot judge whether the framing is appropriate for a specific audience with specific stakes.

The design dimension of visualization, what Cleveland (1993) and Wickham (2016) have systematized as principled approaches to graphical communication, also resists full automation. Choosing the right visual encoding, layering additional variables through color or faceting, and composing graphics that reveal structure rather than obscure it are acts of judgment that require understanding both the data and the audience.

### 2.2.5. GDS5: Data Modeling — Automated for Routine Tasks

Both modeling cultures that Breiman (2001) identified—generative modeling (fitting stochastic models and making inferences about the data-generating mechanism) and predictive modeling— are well within GAI's capability envelope for standard tasks. GAI systems can fit regressions, train random forests, tune hyperparameters via cross-validation, evaluate predictive performance, and interpret coefficients, all from natural-language prompts.

The convergence observation from Section 2.1 applies with particular force here. For a fixed dataset and a well-specified modeling task, different GAI systems will converge on essentially the same fitted model. A linear regression is a linear regression; the coefficients do not depend on which system wrote the code. Even for more complex models, the convergence in approach means that different systems will tend to select similar algorithms, similar preprocessing steps, and similar evaluation metrics for standard problems.

The human judgment that resists automation lies elsewhere: in choosing between generative and predictive frameworks (and then which models), in specifying causal assumptions and assessing whether they are defensible, in recognizing model misspecification, in understanding what a model's outputs mean substantively, and in deciding what to do when diagnostics reveal problems. These are the activities that constitute statistical and computational reasoning, not the execution of a model-fitting pipeline, but the design and interrogation of that pipeline in light of the scientific question and the data-generating process. Breiman's central insight was that the choice between modeling cultures is itself a judgment call, one that depends on whether the goal is explanation or prediction, on what the data can support, and on what the stakeholders need. This choice is a human core competency that GAI cannot make, because it requires understanding the purpose of the analysis, not just its mechanics.

### 2.2.6. GDS6: Science About Data Science — More Important Than Ever

Donoho's sixth division is his most forward-looking: the systematic, evidence-based study of how data analysis is practiced and how it might be improved. He envisioned a field that would study commonly occurring workflows, measure the effectiveness of standard analytical procedures, and uncover emergent patterns and artifacts in published results. He pointed to meta-analysis of the scientific literature (Ioannidis, 2005), cross-study validation exercises (Bernau et al., 2014), and cross-workflow analyses (Madigan et al., 2014) as early examples of this science in action. GAI makes GDS6 more important, not less. As automated analysis becomes widespread, by prompting language models rather than by hand-coding pipelines, the need for systematic evaluation of what these pipelines produce grows correspondingly. Who is checking the GAI-generated analysis? How do we know it is correct? How do we evaluate whether one analytical workflow produces more valid conclusions than another across a body of studies?

The question of reproducibility is particularly interesting in this context. In one sense, GAI may aid reproducibility: generated code can be logged, prompts can be recorded, and entire analytical workflows can be documented with less friction than traditional hand-coded analyses. In another sense, GAI introduces new sources of variation. Language model outputs can vary across sessions, across model versions, and across providers. The RLHF convergence noted in Section 2.1 mitigates this somewhat (different systems tend to produce similar outputs for standard tasks), but version updates, stochastic sampling, and prompt sensitivity mean that exact reproducibility is not guaranteed. Understanding and managing this variation is itself a GDS6 problem.

Checking the results of GAI-generated analyses (code, statistical output, interpretations) should follow the same principles as human review of another human's work. Code review is a well-established practice in software engineering, and its logic extends naturally to GAI outputs: does the code do what it claims? Are edge cases handled? Do diagnostics look reasonable? Are assumptions met?

Looking further ahead, one can envision multi-agent review workflows in which task-specific AI systems review different components of a data science analysis, with one checking statistical validity, another auditing fairness, another verifying reproducibility, before a human orchestrator synthesizes the assessments and makes final judgments. This is GDS6 operationalized: a structured, evidence-based approach to evaluating analytical pipelines, with automation assisting the evaluation rather than replacing it.

## 2.3. The Hierarchy Becomes Inescapable

Table 1 summarizes the division-by-division analysis. The pattern is clear: the divisions most transformed by GAI are those focused on technical execution, while the divisions most resistant to automation are those requiring judgment, context, and meta-reasoning.

**Table 1. GAI's impact on the six divisions of Greater Data Science.** Each division is assessed for what GAI can automate, which human core competencies it requires, and its overall resilience to automation.

| GDS Division | What GAI Automates | Human Core Competency Required | Resilience |
|---|---|---|---|
| GDS1: Data Gathering, Preparation, and Exploration | Cleaning scripts, anomaly flagging, exploration suggestions | Problem formulation, measurement and design, domain knowledge | High |
| GDS2: Data Representation and Transformation | Reshaping, format conversion, SQL, ETL pipelines | Judgment about appropriate representations | Moderate |
| GDS3: Computing with Data | Code in multiple languages, packages, debugging, reports | Computational reasoning (exercised through evaluation, not coding) | Low |
| GDS4: Data Visualization and Presentation | Standard plots, dashboards, chart generation | Sensemaking: audience, framing, uncertainty communication | Moderate |
| GDS5: Data Modeling | Model fitting, tuning, cross-validation, interpretation | Statistical reasoning, causal identification, framework choice | Low (routine) to High (non-routine) |
| GDS6: Science about Data Science | Logging workflows, recording prompts | Reproducibility evaluation, workflow assessment, meta-analysis | High |

Empirical evaluations confirm this pattern. Evkaya and Carvalho (2026), writing in this journal, systematically assessed ChatGPT's Data Analysis plugin across exploratory analysis, visualization, supervised learning, and unsupervised learning tasks. The tool performed competently on well-specified computational tasks (loading data, fitting standard models, producing basic plots) but yielded mediocre results overall. The failure modes clustered around judgment: visualizations required human fine-tuning because the tool could not assess readability for a specific audience; unclear prompts produced poor analyses because the tool could not infer what the analyst actually needed; and the authors concluded that "sustained expert oversight" was essential rather than optional. In the terms of this paper, ChatGPT DA succeeded at GDS3 and routine GDS5 but struggled at GDS1, GDS4, and GDS6, precisely the divisions where the human core is essential for high quality analysis (the Appendix provides a further illustration using a classic econometric dataset).

Donoho's hierarchy was already right in 2017: GDS1 matters more than GDS5, as measured by practitioner time, and GDS6 represents the field's most important intellectual frontier. GAI makes this hierarchy inescapable. When the data science stack is absorbed by language models, educators and students can no longer defer the hard work of judgment by staying busy with code. The human core—problem formulation, measurement and design, causal identification, statistical reasoning, ethics and accountability, sensemaking—is not a residual category left over after automation. It is the substance of what data science was always supposed to be about. The next section unpacks this core in detail.

## 3. Implications for Data Science Curriculum and Assessment

Section 2 established that GAI transforms the six divisions of Greater Data Science unevenly: technical execution is substantially automated, while judgment-laden activities remain essentially human. The practical question is what this means for designing learning outcomes, curricula, and assessments in data science education. This section presents the educational implications in four parts: focus on the human core, integrating GAI (POP cycles and structured review), why sustained engagement with an application area is essential, and how assessment should change.

### 3.1. The Human Core: What Students Must Learn

GAI strips away much of the data science stack that often surrounds (and sometimes obscures) competencies requiring judgment. These competencies overlap substantially with longstanding goals in statistics and data science education (ASA, 2014; National Academies, 2018; De Veaux et al., 2017; Tu et al., 2024). I organize them into six elements.

*Problem formulation* is the act of translating a vague question into a precise analytical question that data can address by specifying an estimand, identifying the relevant population, and determining what comparisons matter. Consider a public health agency that asks, "Is this intervention working?" Before any data is collected, the data scientist must decide: working for whom? Compared to what? Over what time horizon? GAI can suggest possible formalizations, but it cannot judge whether a particular formalization captures what the stakeholder actually needs to know. Problem formulation is the competency most dependent on domain knowledge, which is why sustained engagement with an application area is essential (see Section 3.3).

*Measurement and design* involve choosing and justifying a design for data collection (e.g., experiment or observational study) and understanding the measurement instruments that will produce the data. Design decisions are irreducibly contextual: whether a randomized experiment is feasible depends on ethical, logistical, and institutional constraints that no language model can currently assess from a prompt. Much of the data that industrial data scientists encounter was not collected for the purpose to which it is now being put. Behavioral data generated by commercial platforms are often collected to serve the commercial logic described in Section 1.3, not to answer research questions. Using such data for inference requires understanding the gap between the data-generating process and the research question.

*Causal identification* requires structural reasoning that goes beyond model fitting including specifying assumptions under which a statistical quantity can be interpreted as a causal effect and defending those assumptions with substantive knowledge. This is where GAI's convergence on conventional approaches can generate misleading output. A language model prompted to analyze an observational dataset may produce a regression or matching analysis. If the underlying causal structure involves unmeasured confounders or collider bias, the conventional pipeline can produce misleading conclusions that look entirely plausible. Detecting this failure requires reasoning that connects statistical concepts to specific data-generating mechanisms, the kind of reasoning that distinguishes students who can identify a concept from those who can trace its consequences.

*Statistical and computational reasoning* is the capacity to understand why methods work, not merely how to execute them. It includes reasoning about uncertainty, model diagnostics, sensitivity analysis, and computational validity. If the RLHF convergence conjecture (Section 2.1) is correct, then GAI converges on conventional approaches and the distinctive value of a human data scientist lies in recognizing when the conventional approach fails: heteroskedastic errors invalidating standard errors, non-independence inflating cross-validation estimates, post-hoc test selection distorting p-values. Computational reasoning, as Nolan and Temple Lang (2010) emphasized, is about understanding what computations do and how to validate results, not about memorizing syntax. In the GAI medium, this reasoning is exercised through evaluating outputs rather than writing code, but the intellectual work is the same.

*Ethics and accountability* are structural competencies that pervade the entire analytical workflow, not standalone modules. Every stage of a data science project involves choices with ethical dimensions such as, what data to collect, whose interests the analysis serves, and consequences of errors. Every stage requires accountability practices that make those choices transparent: documenting data provenance, recording analytical decisions, ensuring reproducibility, and producing work that can withstand external scrutiny. The surveillance capitalism context introduced in Section 1.3 makes this especially salient: much of the data that commercial data scientists work with was generated by systems designed to translate human experience into prediction products with material consequences for individuals. These competencies are medium-invariant. They mattered when analyses were coded by hand, and they matter more now that GAI enables analyses to be produced at scale with minimal human oversight.

*Sensemaking* is the activity of translating analytical results into understanding that is actionable by specific audiences. It encompasses not just communication in the narrow sense—clear charts, well-written reports—but the broader work of identifying who the stakeholders are and what decisions the analysis informs, framing uncertainty in terms non-technical audiences can reason with, anticipating misinterpretation, and engaging decision-makers in jointly interpreting findings rather than passively receiving them. Sensemaking is inherently relational: it requires understanding the institutional environment in which results will land, which in turn requires domain knowledge, a fluency that GAI, lacking knowledge of the specific audience and stakes, cannot supply.

These six competencies are not independent. Sound causal reasoning depends on problem formulation, measurement and design, and domain knowledge. Sensemaking requires statistical reasoning to frame uncertainty accurately. Ethics and accountability apply at every stage. The human core is a web of interconnected capacities, not a checklist.

## 3.2. Working with GAI: POP Cycles and Structured Review

The human core tells us *what* students must learn. The question remains *how* they exercise these competencies when GAI is the medium through which they work.
Working with GAI in data science is not a single act of prompting. Empirical studies of LLM-assisted data analysis consistently find that single-turn prompts are insufficient for most realistic tasks: they may produce code that might run but often addresses the wrong question, omit critical preprocessing steps, or generate plausible-sounding but incorrect interpretations (Liu et al., 2024). Data analysis is inherently multi-step and context-dependent where each decision (how to handle missing values, which variables to include, what diagnostics to run) depends on what was learned in prior steps. This sequential, judgment-laden structure resists compression into a single prompt.

Effective use of GAI for data analysis is therefore an iterative process I call the POP cycle: prompt, output, prompt. The practitioner issues a prompt requesting, say, an exploratory analysis of a dataset, examines the output, and then issues a refined prompt based on what the output reveals: asking for diagnostics, testing an edge case, modifying an assumption, or probing an unexpected result. (see Appendix for an illustration of the POP cycle) Each cycle requires human judgment. Is the output correct? Does it address the right question? What should I check next? Critically, POP cycles typically involve retrieval-augmented generation (RAG): the practitioner uploads the dataset, existing code, documentation, or prior analyses into the conversation so that the model operates on the specific analytical context rather than generating from general knowledge alone. This grounding in actual data and project artifacts is what makes the POP cycle a genuine analytical practice rather than an abstract exercise in prompt engineering.

The POP cycle is the vehicle through which the human core competencies are exercised in the GAI medium. Problem formulation shapes the initial prompt and its successive refinements. Statistical reasoning governs the choice of diagnostics. Causal thinking determines whether a suggested analysis is appropriate. Sensemaking guides how the final output is framed. Without these competencies, the POP cycle degenerates into a shallow loop—accepting the first output, or prompting without knowing what to look for—which is the novice dilemma identified in Section 1. Before GAI, students learned to reason about data by writing and revising code. In the GAI medium, the POP cycle is the analogous vehicle. The cognitive work is not less demanding;

if anything, it requires more conceptual understanding, because the student must evaluate outputs they did not construct line by line.

Checking the results of a GAI-generated analysis (code, statistical output, interpretation) should follow the same principles as reviewing another human's work. Code review is a well-established practice in software engineering: a second person reads the code, checks that it does what it claims, tests edge cases, and verifies that the logic is sound. Data analysis or statistical analysis plans are often mandatory parts of biomedical research and require statistical review. The same discipline applies to GAI outputs. Framing GAI review this way couches this type of review in longstanding practices meant to support high quality analyses. Students would not be performing some exotic new task called "AI auditing." They are doing what competent professionals have always done when receiving analytical work from a collaborator — reading it critically, testing it, and taking responsibility for the final product.

Looking further ahead, specialized AI systems may review different components of an analysis before a human makes final judgments: one checking statistical validity, another auditing for fairness, another verifying reproducibility. In this world the human's role could shift from sole reviewer to orchestrator by synthesizing automated assessments, resolving conflicts, and making the judgment calls that require contextual understanding. This connects directly to Donoho's GDS6: a structured, partially automated approach to evaluating analytical pipelines, with the human orchestrator asking whether the analysis is valid and whether the conclusions are warranted.

## 3.3. Application Area Engagement

The human core competencies are not abstract. They are always exercised *within a domain*. Problem formulation requires domain knowledge. Measurement and design require understanding field-specific conventions. Sensemaking requires knowing who the stakeholders are and what decisions the analysis is meant to inform. A data science program designed without sustained engagement in an application area may produce students who can operate tools but cannot exercise judgment. This is not a new argument. Nolan and Temple Lang (2010) structured their courses around substantive projects—geo-location using wireless signals, spam filtering, elephant seal migration—precisely because computational skills need context to become meaningful. Donoho (2017) recommended baseball analytics as a vehicle for developing GDS1 skills. Tukey (1962) argued for apprenticeship with real data in real scientific settings.
Data science curricula should therefore include sustained study of an application area in the sciences, social sciences, or humanities. This means more than analyzing a dataset from an unfamiliar field as a homework exercise. It means enough engagement with the application domain that students can appreciate how questions are formulated that matter in that domain, understand the measurement conventions and data-generating processes that produce the data, assess whether analytical results make substantive sense, and communicate findings in the domain's language.

## 3.4. Assessment That Evaluates Reasoning

If the human core is the curriculum, then assessment must evaluate whether students have developed the competencies that constitute it. Lopez-Miranda et al. (2026), in a qualitative study of instructor perspectives published in this journal, report that data science educators are already grappling with this challenge: tasks once used to teach coding, writing, and problem-solving can now be completed by LLMs, forcing a reconsideration of both pedagogy and assessment. This requires a shift away from assessments that reward correct final answers—which GAI can now produce for most standard problems—toward assessments that evaluate reasoning, design, and judgment.

Concrete formats suited to this goal might include design critiques (given a proposed study, identify threats to validity and suggest improvements), audit reports (given a completed analysis, assess its assumptions, document its provenance, and evaluate its fairness), POP cycle documentation (submit the full sequence of prompts, outputs, and refinements that produced an analysis, annotated with the reasoning behind each iteration), context instructions and rules for configuring an AI agent, and stakeholder communication exercises (translate an analysis into a memo for a non-technical decision-maker, framing uncertainty and limitations honestly). These formats assess different competencies such as, design critiques target measurement and causal identification; audit reports target ethics and accountability; POP documentation targets statistical reasoning and computational thinking; communication exercises target sensemaking, but they share a common structure: they require students to show their reasoning, not just their results. The Appendix provides an example of what such annotated documentation might look like in practice.

Assessment should distinguish students who merely identify a concept from those who connect it to mechanisms and human consequences. A student who writes "there may be confounding" has identified a concept. A student who specifies which variables might confound the analysis, explains the mechanism through which they operate, and traces the direction and magnitude of the resulting bias has demonstrated causal reasoning. The difference matters because current GAI can produce the first response easily; only a student with genuine understanding can produce the second.

GAI could become a medium for deeper learning and trustworthy decisions rather than a shortcut to shallow answers. The challenge is to ensure that students develop enough prior knowledge to resist the medium's distortions: to catch the equivalent of 12×17 = 29, as introduced in Section 1.

## 4. Conclusion

Generative AI rests on general-purpose methods such as search and statistical learning that scale with computation (Sutton, 2019) rather than on transparent, domain-specific models. These methods are powerful enough to automate much of the routine data science workflow embedded in curricula developed in the last decade. But the intellectual tradition from Tukey through Donoho has long argued that this workflow, however necessary, was never the field's most important contribution. The most important work was always the judgment-laden, context-dependent, human-facing activity of formulating questions, designing studies, reasoning about evidence, and making sense of results for the people who need them. GAI makes this hierarchy inescapable. Future data scientists will need to pose questions, design studies, interrogate models, and take responsibility for the consequences of automated pipelines. Understanding the commercial and institutional contexts that produce data, working effectively within a medium capable of generating plausible but potentially misleading outputs at scale, and exercising judgment about when conventional approaches fail are not skills that automation renders obsolete.

Data scientists are uniquely positioned among the professions being reshaped by GAI. The statistical and computational foundations of these systems—gradient descent, regularization, cross-validation, reinforcement learning, bias-variance tradeoff—are part of the discipline's own toolkit. Where lawyers, physicians, or educators encounter AI as an external disruption, many in the data science community have an insider's understanding of how these systems work, not just what they produce. As Lin et al. (2025) have argued, statisticians and data scientists can be leaders rather than merely collaborators in AI research, precisely because AI tools are essentially algorithms whose behavior can be studied using the discipline's own methods. The educational implication is that data science curricula should equip students not only to use GAI but to understand its behavior from the inside and reason about why outputs converge, where hallucination comes from, and how to evaluate model reliability. This is an advantage no other field can claim, and it should be cultivated deliberately.

Whether the human core competencies are permanently irreducible, reflecting the inherent structure of learning from data about the world, or merely beyond the reach of current AI systems, the curricular implications are the same: students who develop genuine reasoning about data, evidence, and consequences will be prepared for any future, including one in which AI systems eventually share these capacities. The medium has changed but the human core endures. Defining learning outcomes and curriculum around these competencies, aligning instruction with the new medium through which students will practice, and assessing reasoning rather than answers is the work of data science education in the generative AI era.

**Disclosure Statement**

Nathan Taback has no financial or nonfinancial disclosures to share for this article.

Wu, C. F. J. (1997). *Statistics = Data Science?* Inaugural Carver Lecture, University of Michigan. https://www.isye.gatech.edu/~jeffwu/presentations/datascience.pdf

Zhuo, T. Y., Vu, M. C., Chim, J., Hu, H., Yu, W., Widyasari, R., … & Ong, D. (2024). BigCodeBench: Benchmarking code generation with diverse function calls and complex instructions. *arXiv preprint arXiv:2406.15877*. https://doi.org/10.48550/arXiv.2406.15877

Zuboff, S. (2019). *The Age of Surveillance Capitalism: The Fight for a Human Future at the New Frontier of Power*. PublicAffairs.

**Appendix: A POP Cycle in Action**

The following example illustrates how a POP cycle exercises the human core competencies described in Section 3. The example uses the Engel (1857) food expenditure dataset, a public-domain dataset of 235 Belgian working class households with two variables: annual household income and annual food expenditure (in Belgian francs). The dataset is distributed with the quantreg package in R and the statsmodels library in Python (Koenker & Bassett, 1982). It was chosen because heteroscedasticity is a well-documented feature of Engel curve estimation, though this fact was not mentioned in any of the prompts.

The transcript below reproduces an actual conversation with Claude Opus 4.6 Extended (Anthropic, March 2026). The same prompts were also issued to ChatGPT 5.4 Thinking (OpenAI), which produced nearly identical results at each stage, consistent with the RLHF convergence discussed in Section 2.1. Annotations in italics identify where human core competencies intervene.

**Cycle 1: Initial prompt and output**

*Prompt:* "I've uploaded a CSV file with household survey data. The columns include household_income and food_expenditure. Fit a linear regression predicting food expenditure from income. Report the coefficients, standard errors, p-values, and R-squared."

*GAI output (abridged):*
Model: foodexp = 147.48 + 0.485 × income

|  | Coefficient | Std. Error | t-statistic | p-value |
|---|---|---|---|---|
| Intercept | 147.48 | 15.96 | 9.24 | $1.57 \times 10^{-17}$ |
| Income | 0.485 | 0.014 | 33.77 | $9.92 \times 10^{-92}$ |

$R^2 = 0.830$, Residual SE = 114.11
Interpretation: Both coefficients are highly significant. The slope of ~0.485 means that for every additional dollar of household income, food expenditure increases by about 49 cents. The model explains roughly 83% of the variation in food expenditure, indicating a strong linear relationship. This is consistent with Engel's Law: food expenditure rises with income, but less than proportionally. Let me know if you'd like diagnostics (residual plots, normality tests, etc.).

*What a novice might do:* Accept the output. The $R^2$ is high, the p-values are tiny, and the interpretation sounds authoritative. Copy it into a report.

*What the human core demands:* The data scientist notices that the GAI offered diagnostics but did not flag any concerns about the model. The results are presented as clean. But the data scientist recalls that expenditure data often exhibits increasing variance with income: wealthier households have more discretion in spending, producing a fan-shaped pattern in residuals. This suspicion comes from domain knowledge (problem formulation) and statistical reasoning (awareness that OLS standard errors assume constant variance). The GAI did not raise this concern. The data scientist must know to ask.

### Cycle 2: Diagnostic prompt

*Prompt:* "Plot the residuals against fitted values from this regression. Also plot the residuals against income. I want to check for heteroscedasticity. Run a Breusch-Pagan test as well."

*GAI output (abridged):*
**Breusch-Pagan test:** LM statistic = 109.26, df = 1, p ≈ 0.0000, overwhelmingly rejecting the null of homoscedasticity.
Both plots tell the same story. The residual spread fans out as income increases: low-income households cluster tightly around the regression line, while high-income households show much larger deviations in both directions. This is a textbook pattern for expenditure data. Lower-income households have less discretion in food spending (most income goes to necessities), so their expenditure is more predictable. Higher-income households have more choice, some spend heavily on food, others redirect income elsewhere, producing greater variance.
The OLS coefficient estimates (0.485, 147.48) are still unbiased and consistent. But the standard errors and p-values are unreliable: they will tend to be too small, making significance tests overly optimistic.

*What the human core demands:* The data scientist interprets the diagnostics. The fan-shaped residual plot is a textbook signature of heteroscedasticity, and the Breusch-Pagan statistic of 109.26 is enormous. Notably, the GAI now explains the pattern competently and even identifies the economic mechanism (income discretion). But this explanation was available only because the human asked for it. In Cycle 1, the GAI presented the OLS results without concern. The standard errors of 0.014 (income) and 15.96 (intercept) are too small, inflating t-statistics and producing misleadingly narrow confidence intervals. The coefficient estimates are not biased (OLS coefficients remain unbiased under heteroscedasticity), but the reported precision is overstated. This reasoning connects statistical theory (properties of OLS under violated assumptions) to practical consequences (misleading inference).

### Cycle 3: Corrective prompt

*Prompt:* "Re-estimate the regression using heteroscedasticity-consistent (HC3) standard errors. Compare the original and robust standard errors side by side. Also fit a weighted least squares model using the inverse of predicted variance as weights, and compare the results."

*GAI output (abridged):*
**HC3 robust standard errors vs. classical:**

|           | Classical SE | HC3 SE | Classical t | HC3 t |
|-----------|--------------|--------|-------------|-------|
| Intercept | 15.96        | 59.96  | 9.24        | 2.46  |
| Income    | 0.014        | 0.066  | 33.77       | 7.31  |

The HC3 correction reveals how badly the classical SEs understate uncertainty. The income SE increases from 0.014 to 0.066, over 4.6 times larger. The t-statistic drops from 33.8 to 7.3, still highly significant, but the classical SEs were wildly overconfident. The intercept drops from t = 9.2 to t = 2.5, barely significant at 5%.

**Weighted least squares:**
WLS shifts the intercept from 147.48 (OLS) to 66.43, and the slope from 0.485 to 0.577. Once high-variance observations are properly downweighted, the estimated marginal propensity to spend on food is actually higher than OLS indicated, and the baseline expenditure is lower.
*What the human core demands:* The data scientist now has a defensible analysis, but also a substantively different story. The OLS standard errors were underestimated by a factor of approximately four. More importantly, the WLS estimates tell a different narrative: the relationship between income and food expenditure is steeper than OLS suggested (0.577 vs. 0.485), and the baseline subsistence level is lower (66 vs. 147 francs). These are not minor technical corrections; they change the economic interpretation. Communicating these differences honestly to a policy audience, explaining what the correction means and why it matters, is an act of sensemaking that requires understanding both the statistics and the domain.

**What this example illustrates**
Without the human core, the POP cycle would have stopped at Cycle 1. The GAI executed each computational request competently and even offered to run diagnostics, but it did not identify heteroscedasticity as a concern or suggest that the reported standard errors might be unreliable. The human data scientist brought statistical reasoning (knowing to check residuals and why), domain knowledge (expecting heteroscedasticity in expenditure data), and sensemaking (interpreting the practical implications of different estimates for a policy audience). The fact that both Claude and ChatGPT produced essentially the same results at each stage underscores the convergence discussed in Section 2.1: different systems execute the conventional pipeline reliably, but recognizing when that pipeline is inadequate remains a human responsibility.
The full conversation transcript is available at https://claude.ai/share/e8f82216-7e4e-46cc-8610-bd6d47c28774 .

**Appendix References**

Engel, E. (1857). Die Productions- und Consumtionsverhältnisse des Königreichs Sachsen. *Zeitschrift des Statistischen Bureaus des Königlich Sächsischen Ministeriums des Innern, 8*, 1–54.

Koenker, R., & Bassett, G. (1982). Robust tests for heteroscedasticity based on regression quantiles. *Econometrica, 50*(1), 43–61. https://doi.org/10.2307/1912528